\documentclass[twocolumn, secnumarabic, amssymb, nobibnotes, aps, prl]{revtex4-2}

\usepackage{array}
\usepackage{soul,color}
\sethlcolor{yellow}

\soulregister\cite7  
\soulregister\ref7
\soulregister\pageref7

\usepackage{xcolor}
\usepackage{graphicx}
\usepackage[detect-all]{siunitx}
\usepackage{amsmath}
\usepackage{amssymb}                                                               
\usepackage{array}                                            
\usepackage{IEEEtrantools}
\usepackage{url}                          

\graphicspath{ {./Figuresnew/}}

\begin{document}

\title{Short- and long-term avalanche dynamics in 1D-printed microfluidic crystals}%
\author{Ronald \surname{Terrazas Mallea}}
\email{rtmallea@ichf.edu.pl} 
\author{Jonathan \surname{Pullas Navarrete}}
\author{Jan \surname{Guzowski}}
\email{jguzowski@ichf.edu.pl}
\affiliation{Instytut Chemii Fizycznej PAN, ul. Kasprzaka 44/52, 01-224 Warsaw, Poland}

\begin{abstract}
\noindent
We report arrays of monodisperse water-in-oil microdroplets printed onto a substrate in a form of a compact linear chain---a 1D-crystal---pinned at one end. The chain spontaneously collapses under capillary forces via a sequence of avalanche-like rearrangement waves, resembling the rearrangements in a flowing microfluidic crystal, yet  limited by the hydrodynamic friction at the substrate. While the propagation of the subsequent waves, separated by highly ordered metastable states, is either accelerating or decelerating depending on the direction of collapse, the coarse-grained dynamics of multiple waves---at moderate packing fractions $\phi$---is initially linear in time, before leveling off. We further demonstrate how the collapse  can be prevented via the use of a roughened substrate. Our study provides insight into the short- and long-term avalanche dynamics in granular systems with free interfaces and opens way to precision-printing of microfluidic  assays.         
\end{abstract}

\date{January 2023}
\maketitle 

Soft granular materials, consisting of  close-packed deformable grains separated by thin membranes or lubricating fluid films, are ubiquitous in biology, food or pharmaceutical industry. Examples  range from foams \cite{hohler2005RheologyLiquidFoam}, concentrated emulsions \cite{macminn2015FluidDrivenDeformationSoft} and microgel suspensions \cite{daly2020HydrogelMicroparticlesBiomedical} to biological tissues \cite{astrom2006CellAggregationPacking, smeets2020CompactionDynamicsProgenitor, wyatt2015EmergenceHomeostaticEpithelial}. Processing of such materials  at ever-smaller scales---down to the single grain level---has been of increasing interest in  tissue-engineering \cite{daly2020HydrogelMicroparticlesBiomedical, xie2022Situ3DBioprinting}. Since the typical size of  the grains  (100-1000 $\mu$m) significantly exceeds the size of a living cell (5-10 $\mu$m),  each grain can conveniently serve as a capsule or a biological niche. In particular, extrusion-printing of soft-granular materials remains one of the basic strategies in fabrication of porous or cell-laden granular scaffolds, e.g., for bone- or vascular tissue engineering \cite{VisserSciAdv2018,CostantiniGuzowskiAngewChem2019, daly2020HydrogelMicroparticlesBiomedical, SchotLeijtenAdvMat2024}. 

In general, the printability of soft granular  materials is directly related to their  rheological behavior \cite{HighleyBurdickAdvSci2019}. Combination of shear-thinning and a finite yield-stress allows for efficient extrusion of stable threads. However, in the case of ultrathin  threads, one- or few grains in diameter, the  capillary forces acting at the interface \cite{guzowski2022DynamicSelforganizationAvalanching} as well as local ordering effects \cite{Ono-Dit-BiotDalnoki-VeressSoftMatter2021} become the dominant factors. One can expect that such thin granular structures should behave similarly  to the so-called microfluidic crystals, conventionally generated inside microchannels and carried by  external flows \cite{DelGiudiceMaffettoneLabChip2021}. The structural instabilities in the flowing droplet or bubble crystals are known to proceed 
via   \emph{rearrangement waves}  and transitions between different ordered states \cite{RavenMarmottantPRL2009,guzowski2022DynamicSelforganizationAvalanching}. However,  the  relaxation dynamics of the \emph{printed} soft-granular crystals with free interfaces has not been  studied previously. Besides  the basic interest in soft matter physics, the problem remains of exceptional relevance in precision-bioprinting and applications such as high-throughput drug testing exploiting encapsulated '3D' cell-cultures  or in microtissue/organoid engineering \cite{ZhouBayleyAdvMat2020}.

In this Letter, we report microfluidic generation of close-packed water-in-oil emulsion droplets and their direct printing one-by-one at a substrate under an external immiscible aqueous phase in a form of a linear chain (\textbf{Fig.\ 1a,b}, movies \textbf{M1, M1b}). Our focus is  the relaxation dynamics of such generated quasi-1D droplet crystals.  Within short time after printing, the chain--- pinned at one end---becomes unstable and gradually collapses via a sequence of rearrangement waves  (\textbf{Fig.\ 1b}). We  find that the collapse of a single wave is either decelerating or accelerating depending on the direction of collapse, while
the long-term dynamics of multiple waves---at moderate packing fractions, $\phi\sim 0.75$ (we define $\phi=Q_w/(Q_o+Q_w)$ where $Q_w$, $Q_o$ are the rates of flow of water and oil, respectively)---is initially linear in time, before eventually leveling off. We propose a model explaining both the short- and long-term behavior in this regime.  At high $\phi$ ($\phi>0.8$) the occurrence of the waves becomes  stochastic and dominated by the inter-avalanche waiting times. Finally, we show the collapse can be totally prevented via the use of a substrate with micrometric roughness. 

\begin{figure}[t]
\includegraphics[width=\columnwidth]{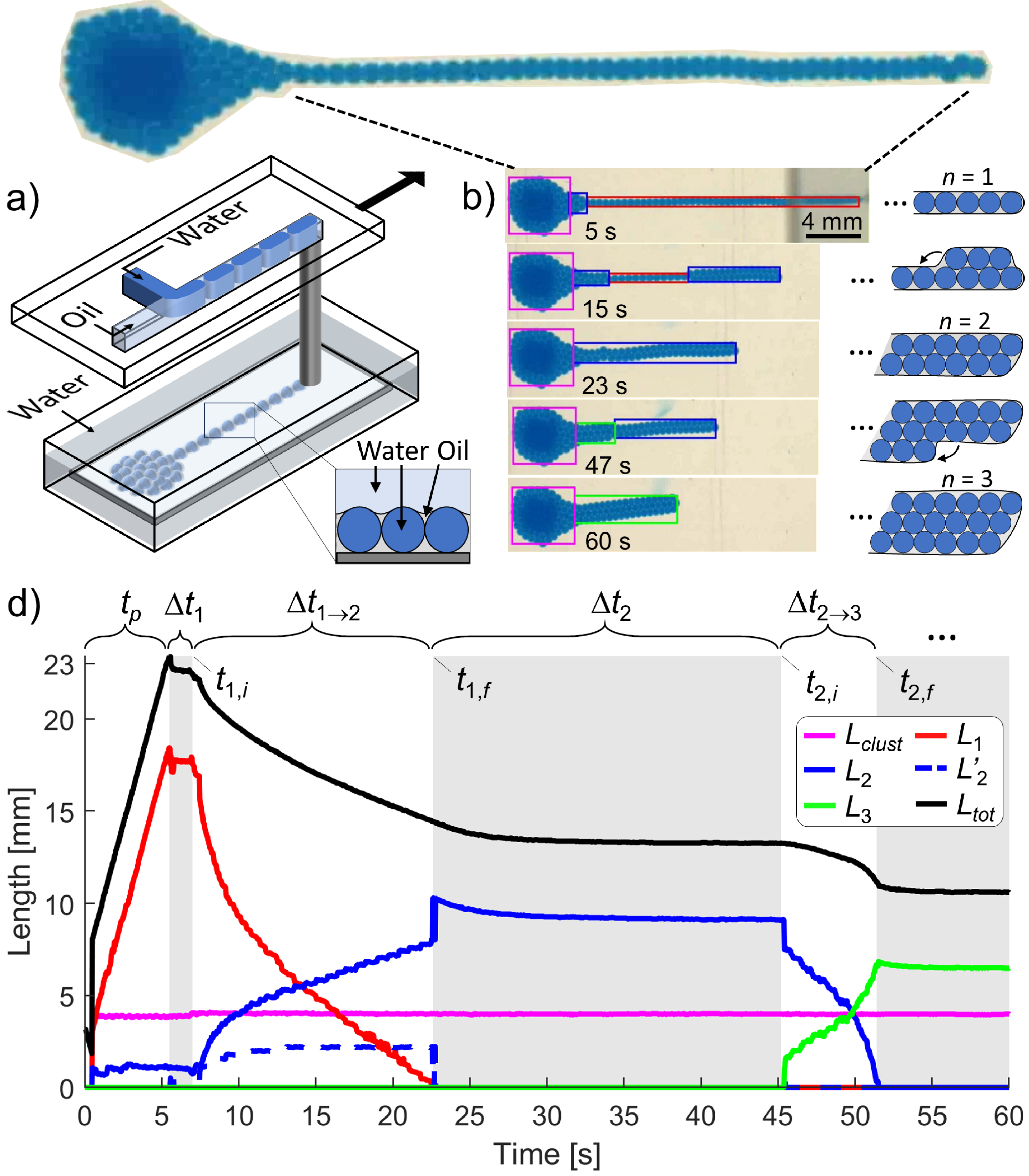}
\caption[Experiment description]{(a) Sketch of a T-junction generating aqueous droplets at a high volume fraction $\phi$ in the carrier oil phase. The droplets are extruded onto a hydrophobic substrate submerged under an external aqueous phase. (b) Time-series of snapshots of the collapsing droplet-thread attached to a large cluster, with indicated bounding boxes around the different detected substructures; $t=0$ corresponds to the beginning of the extrusion process.  (c) The lengths of the substructures  as a function of time. The colors  of the  lines correspond to the colors of the bounding boxes in (b). The dashed blue line corresponds to the $(n=2)$-substructure close to  the cluster (snapshots at $t=5$s, $15$s). The black curve is $L_{tot}$. }
\label{fig:SketchExperiment}
\end{figure}

We use a polycarbonate printhead consisting of a micromilled  T-junction (150$\mu$m$\times$150$\mu$m, rendered hydrophobic using 3M Novec 1720) supplied by a syringe pump (Nemesys S,  Cetoni) to generate the W/O emulsion droplets (diameter $D\sim$ 300$\mu$m) at packing fractions  $\phi\geq0.75$ and directly transfer them onto a substrate  via a blunt outlet  needle (25G), see \textbf{Fig.\ 1a}. The movement of the printhead is provided by  a commercial 3D printer  (Gate 3Novatica). We use water with 0.1 \% w/w Erioglaucine (blue dye, Sigma Aldrich) as the droplet phase and hydrofluoroether fluid Novec 7500 (3M) with 3  \% w/w PFPE-PEG-PFPE fluorosurfactant (Chemipan, Warsaw, PL) as the oil phase. 
We also use water as the external phase. The substrate is made of fluorinated thermoplastic  THV 500GZ (3M Dyneon) \cite{begolo2011NewFamilyFluorinated, nightingale2020EasilyFabricatedMonolithic} cast from a smooth polydimethylsiloxane (PDMS) slab. 
The fluorinated oil phase preferentially wets the fluorophilic surface resulting in  capillary adhesion of the non-wetting droplets and their effective 2D-confinement.  

The printing process starts with a deposition   of the cluster ($N_{clust} \sim 120$ droplets, $t=0$) followed by  extrusion of the single-file structure ($t=t_p$) consisting of  $N \simeq 40$ droplets. 
During collapse, we  distinguish between single-row, double-row, triple-row structures, etc., which we also refer to as the $n$-structures with $n=1,2,3,...$, respectively (\textbf{Fig.\ \ref{fig:SketchExperiment}b}). The  lengths $L_n$ or  $L_{clust}$ are measured from the detected bounding boxes, and the total length  is  $L_{tot}=L_{clust} + \sum_nL_n$.

The collapse proceeds via a sequence of rearrangement waves $W_{1\rightarrow2}$, $W_{2\rightarrow3},...$ whose duration times we denote as $\Delta t_{1\rightarrow2}$, $\Delta t_{2\rightarrow3},...$ (\textbf{Fig.\ 1c}). The times when the $n$-wave starts and ends we denote as  $t_{n,i}$ and $t_{n,f}$, respectively, so that $\Delta t_{n \rightarrow n+1}=t_{n,f}-t_{n,i}$. Between the waves, the system is temporarily trapped in the highly ordered  $n$-states. We denote the  inter-avalanche waiting times as $\Delta t_1,\Delta t_2,...$,
such the whole process can be described as a sequence of intervals  $t_{p},\Delta t_1, \Delta t_{1\rightarrow2}, \Delta t_2, \Delta t_{2\rightarrow3},\Delta t_3,...$. In the experiment, we limit the observation time to  $t_{exp} =  60$ s. In most cases, we observe only  two rearrangement waves, but in some cases ($\phi=0.75$, movie \textbf{M4}),  we also observe third or even fourth wave.      

We find that the first rearrangement wave $W_{1\rightarrow2}$ starts always at the free end and we refer to this type of  collapse as \emph{outside-in}. 
As $(n=1)$-structure collapses (red line in \textbf{Fig.\ 1c}), the $(n=2)$-structure emerges (blue line in \textbf{Fig.\ 1c}), 
and the dynamics is decelerating. The second wave $W_{2\rightarrow3}$ starts  at the pinned end, i.e., at the cluster, and we refer to such wave as \emph{inside-out}. 
Here, the $(n=2)$-structure collapses while the $(n=3)$-structure emerges (green line in \textbf{Fig.\ 1c}), and the dynamics  is accelerating.
Noteworthy, all  the subsequent higher-order waves (if present) are always \emph{inside-out} and have the dynamics qualitatively  similar to the second wave.
 
At $\phi=0.75$ and $\phi=0.833$ we find highly reproducible dynamics $L_2(t)$ throughout multiple experimental runs 
(for $\phi=0.75$ see \textbf{Fig.\ 2a,b}). In further analysis, we focus on the first and second wave and empirically 
find the scalings $L_2(t) \sim (t-t_{1,i})^{\alpha}$ and  $L_2(t) \sim (t_{2,f}-t)^{\beta}$, respectively, 
with $\{\alpha,\beta\}=\{0.46,0.54\}$ for $\phi=0.75$ (solid lines in \textbf{Figs.\ 2a,b}) and $\{\alpha,\beta\}=\{0.42,0.48\}$ for $\phi=0.833$. 

To explain the power-laws we assume that the viscous dissipation occurs mostly within the thin film of oil at the substrate 
whose thickness $h$, based on previous studies \cite{BoreykoCollierPNAS2014}, we may estimate as $h< 10\, \mu$m. 
The Reynolds number in the thin film is Re $=\rho hU/\eta$, where $\eta$ and $\rho$ are the viscosity and the density of oil, 
so that Re $<10^{-5}\times1.5\times10^{-2}/(0.77\times 10^{-6})\approx 0.2$. Accordingly, the dynamics is dominated  by the hydrodynamic friction 
at the substrate. We assume that the rearrangements are driven mostly by the capillary force $F_{cap}= \lambda_n\gamma_{ext}D$  
acting at  the  external water-oil interface at the rearrangement front, where $\gamma_{ext}$ is the interfacial tension 
of the oil-water (external) interface and $\lambda_n$ is a geometrical coefficient. Since $N_{clust}\gg N$, 
we may treat the thread as being effectively pinned at the cluster. 

In the  case of \emph{outside-in} waves $W_{n\rightarrow n+1}$,  only the $(n+1)$-structure moves relative to the substrate. 
The equation of motion of the collapsing $(n+1)$-structure  can be written as $\lambda_n \gamma_{ext}D = N_{n+1}(t)\zeta\eta DU_{n+1}(t)$, 
where $\zeta$ is the single-droplet hydrodynamic friction coefficient,
$N_{n+1}(t)$ is the number of droplets in the $(n+1)$-structure and $U_{n+1}(t)$ their collective velocity. 
From geometry, we have $U_{n+1}(t)=(D/(n^2+n))(\text{d}N_{n+1}/\text{d}t)$ which leads to   
\begin{equation}
L_{n+1}(t)=\frac{D N_{n+1}(t)}{n+1} 
=D\left(\frac{2n}{n+1}\right)^{1/2}
\left(\frac{t-t_{1,i}}{\tau_n}\right)^{1/2},
\label{eq:Ln1}
\end{equation}
where  $\tau_n=\zeta \eta D/(\lambda_n \gamma_{ext})$ can be interpreted as the time of the \emph{first} rearrangement in the wave 
(since\ $L_{n+1}(t=\tau_n)/D=1$). Thus, our model predicts \emph{decelerating} collapse of \emph{outside-in} waves with a power-law 
exponent $\alpha=1/2$, independent of $n$, in good agreement with the experiment. 
 
\begin{figure}[t]
\includegraphics[width=\columnwidth]{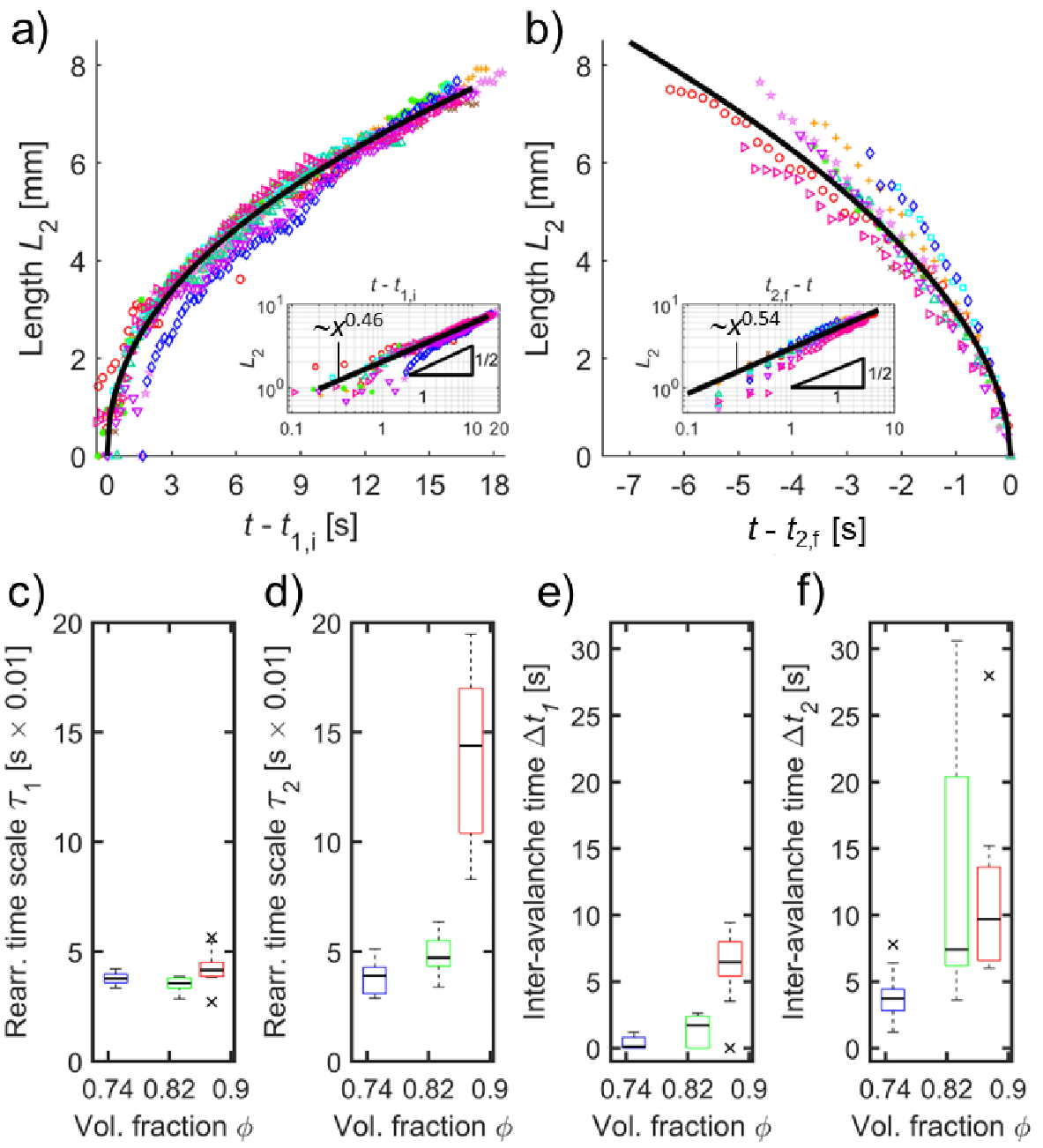}
\caption[Experimental results water]{(a-b) The length of the $(n=2)$-structure $L_2(t)$ (see the blue curve in \textbf{Fig.\ 1c}) 
during (a) the first  and (b) the second  rearrangement wave in several experimental runs at $\phi=0.75$ (symbols) together with 
the fitted power-law trend (black solid curves). Insets show the same data in log-log scale. (c-f) Box-plots of the fitted rearrangement 
timescales $\tau_1$ and $\tau_2$, and the measured inter-avalanche times $\Delta t_1$ and $\Delta t_2$ for  $\phi=\{0.75, 0.833, 0.875\}$ 
(colored blue, green, red, respectively). Horizontal lines are medians; boxes -- 25 to 75 percentile range; 
whiskers -- furthest values not being outliers; $\times$'s -- the outliers, i.e., the points beyond 1.5 times the interquartile range 
away from the bottom or top of the boxes.}
\label{fig:FigDescrip}
\end{figure}

We note that eq.\ (\ref{eq:Ln1}) resembles the so-called Washburn's law \cite{Washburn1921} for the capillarity-driven penetration length $L(t)$ of a viscous liquid into a capillary. In viscous imbibition,  the hydraulic resistance is proportional to $L(t)$, resulting in    $L(t)\sim t^{1/2}$.  In our case, the friction is  proportional to the length of the moving part of the thread $L_{n+1}(t)$ which leads  to the same exponent. 

In the case of the \emph{inside-out}  waves $W_{n\rightarrow n+1}$, the  $n$-structure moves relative to the substrate while  the $(n+1)$-structure is pinned which leads to
\begin{equation}
L_n(t)=D\left(\frac{2n+2}{n}\right)^{1/2}\left(\frac{\Delta t_{n\rightarrow n+1}-t}{\tau_n}\right)^{1/2}.
\label{eq:Ln2}
\end{equation}
Accordingly, we find an \emph{accelerating} collapse of the  \emph{inside-out}   waves with the same exponent,  $\beta=1/2$.
Noteworthy, in both types of waves the $n$'th wave duration time reads 
\begin{equation}
\Delta t_{n \rightarrow n+1}
=\frac{N^2\tau_n}{2n(n+1)}.
\label{eq:Dtnn1}
\end{equation} 
The reproducibility  of the measured rearrangement timescales $\tau_n$ seems  to deteriorate with increasing packing fraction $\phi$ as  the width of the $\tau_n$-distributions increases (\textbf{Fig.\ 2c,d}).
However, at $n=1$, we see little $\phi$-dependence of the median $\tau_n$ (\textbf{Fig.\ 2c}); this means that---somehow counter-intuitively---the substrate friction is nearly  $\phi$-independent. In contrast, at $n=2$ we see a rapid rise of $\tau_n$ at $\phi=0.875$ (\textbf{Fig.\ 2d}). In the second wave  each rearrangement event involves two---rather than one, as in the first wave---topological T1 processes. Accordingly, our results indicate that at  sufficiently high packing fractions the inter-droplet friction starts to impact the dynamics  alongside the substrate friction.

A somehow different trend is observed for  $\Delta t_n$ (\textbf{Figs.\ 2e,f}) which rapidly increases with $\phi$ already at $n=1$. This behavior may reflect the increase in energy barriers for droplet-chain buckling at stronger droplet deformations \cite{guzowski2015DropletClustersExploring}. 

In the following, we make further predictions regarding the long-term dynamics, applicable to the cases with moderate $\phi$ ($\phi\sim 0.75$), in which the stochasticity of $\tau_n$ and $\Delta t_n$ can be neglected. First, we note that the long-term envelope  $L(t)$ approximating  the length of the thread after the $(n-1)$'th rearrangement wave 
 can be cast as $L(t)=ND/n(t)$ where $n(t)$ is found via inverting the total elapsed time $t(n)$, where\begin{equation}
t = \sum_{k=2}^{n} \big[\Delta t_{k-1\rightarrow k}+\Delta t_{k}\big].
\label{eq:t}
\end{equation}
For simplicity, in eq.\ (\ref{eq:t}), we neglect $\Delta t_1$ and set $t=0$ at the onset of the first wave. 

At early stages, that is for  small $n$, i.e., the waiting times  $\Delta t_n$  are typically shorter than wave duration times   $\Delta t_{n\rightarrow n+1}$ (eq.\ (\ref{eq:Dtnn1})), 
so that  
$t\simeq(N^2\tau/2)\sum_{k=2}^{n} (k^2-k)^{-1}=(N^2\tau/2)(1-n^{-1})$. This leads to  
\begin{equation}
L(t)\simeq L_0- \frac{2Dt}{N\tau},
\label{eq:Lt1}
\end{equation}
\noindent where $L_0\equiv L(t=0)=D N$.
Accordingly, at early stages, the envelope $L(t)$ is \emph{linearly} decreasing. The scenario holds until the times $\Delta t_{n\rightarrow n+1}$---which vanish  as $\sim n^{-2}$ (eq.\ (\ref{eq:Dtnn1}))---become  shorter than $\Delta t_n$. At later stages, as the waiting times start to dominate, we may consider two generic cases. 

First, we consider $\Delta t_n =const\equiv \Delta t_0$. This assumption  leads to  
$t=(n-1)\Delta t_0$ and 
\begin{equation}
L(t)\simeq \frac{L_0}{\Delta t_0^{-1}t+1}.
\label{eq:Lt2}
\end{equation}
Second, we consider $\Delta t_n$ diverging at $n=n_{max}$, such that $L(t)\rightarrow L_{\infty}=L_0/n_{max}$. 
As an example, we consider power-law scaling 
\begin{equation}
\Delta t_n =\frac{\Delta t_{max}}{(n_{max} - n)^{\delta}}
\label{eq:Dtn}
\end{equation}
with model parameters $\Delta t_{max}$, $\delta$. Further analysis reveals [\textbf{SI}] that in such a case the dynamics  \emph{qualitatively} depends on  $\delta$. For $\delta<1$, we find that the relaxation takes \emph{finite} time
$t_{max}=\Delta t_{max}(n_{max}-1)^{1-\delta} (1-\delta)^{-1}$.
In contrast, for $\delta>1$, we find an asymptotic decay  \begin{equation}
L(t)-L_\infty \simeq \frac{L_0}{n_{max}^2}\left(\frac{\Delta t_{max}}{(\delta -1)t}\right)^{\frac{1}{\delta-1}}.
\label{eq:Lt3}
\end{equation}
A convincing  experimental verification of these two scenarios would  require printing of extremely long lines, 
presumably with $N\simeq 10^3$, and observation times of the order  $t_{exp}\simeq 10^3$ s (instead of the current $N=40$ 
and $t_{exp}=60$ s), and thus we leave such experiments as a future work. 

\begin{figure}[t]
\includegraphics[width=\columnwidth] {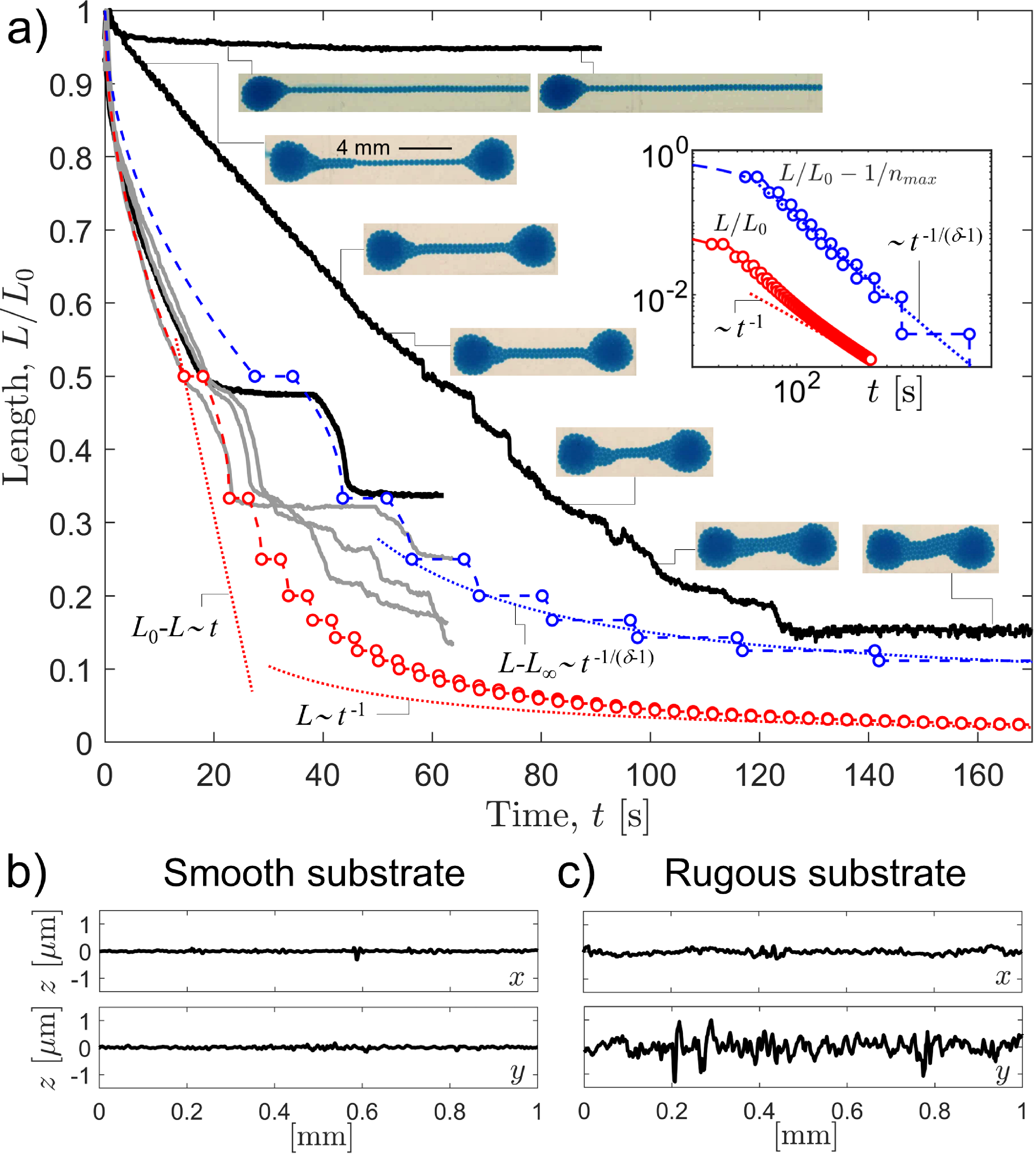}
\caption{(a) Experimental thread length $L=\sum_nL_n$ (cluster excluded) normalized to the initial length $L_0$ in the cluster-thread setup  at $\phi=0.75$ (grey solid curves, movies \textbf{M4-M6}) and at  $\phi=0.833$ (lower black curve, movie \textbf{M1}, \textbf{Fig.\ 1c});  in the cluster-thread-cluster setup (middle black curve, movie \textbf{M2}), and  in the cluster-thread setup at a rugous substrate (upper black curve, movie \textbf{M3}).  For comparison, we show theoretical results for the case with $\tau_n= const=\tau$ and with $\Delta t_n$ either (i) constant (full model---red dashed line---with $N^2\tau=58$s, $\Delta t_n=\Delta t_0=3.5$s; early trend and late asymptote---red dotted lines---eqs.\ (\ref{eq:Lt1}), (\ref{eq:Lt2}))  or (ii) diverging  (full model---blue dashed line---with $N^2\tau=110$s, eq.\ (\ref{eq:Dtn}) with $\Delta t_{max}=440$s,  $n_{max}=13.5$ and $\delta = 1.7$; late asymptote---blue dotted line---eq.\ (\ref{eq:Lt3})). Open symbols mark the onset/end  of each rearrangement waves. Inset shows the  asymptotic behavior of the model (case (i) is multiplied by 0.1 for better visibility). (b-c) Surface profiles of the THV substrates in the directions parallel ($x$) and perpendicular ($y$) to the prints.      }
\label{fig:FigClust2rug}
\end{figure}  

Finally, in the quest for fully stable (non-rearranging) threads, we explore  possible methods of  slowing down or even preventing the collapse. 
First, we pin both ends of the line via printing of another cluster at the other end of the thread.
Instead of the step-wise accelerating/decelerating collapse, we observe approximately linear dynamics (\textbf{Fig.\ 3a}, movie \textbf{M2}). Due to the symmetry, both clusters remain in relative motion with respect to the substrate. Accordingly, the net friction is mostly  provided by the two  clusters, and so remains nearly time-independent. 

In another experiment, we use a \emph{rugous} THV substrate obtained via THV molding from a milled aluminium surface (rather than from the smooth PDMS) and  consisting of parallel micrometric grooves (\textbf{Fig.\ 3b,c}). The  $n=1$ thread,  printed in the direction perpendicular to the grooves and pinned at one end, remains stable for at least several minutes, of which we directly record   $t_{exp}= 90$ s (\textbf{Fig.\ 3a}, movie \textbf{M3}).  We suppose that the grooves support  capillary suction of the oil phase from between the droplets effectively raising $\phi$  and enhancing the capillary arrest.

\vspace*{0.1cm}

\noindent \emph{Discussion}--In summary, we study the collapse of 1D-printed microfluidic crystal: an ordered single-file  thread of close-packed aqueous droplets bound by capillary bridges of an engulfing oil phase. The dynamics of collapse  consists of a sequence of avalanche-like rearrangement waves separated by  ordered metastable states. We  develop a model explaining short- and long-term  dynamics and propose a method of full stabilization of the printed threads via the use of rugous substrates.

Noteworthy, in   our printed granular threads we never observe  Rayleigh-Plateau instability \cite{diez2009BreakupFluidRivulets}, i.e., no spontaneous breakup  into separated   'drops' or sub-clusters. This may be attributed to the strongly adhesive  capillary bridges which lead to effective   elasticity of the structure which therefore resembles a solid fiber  rather than a viscous thread \cite{guzowski2022DynamicSelforganizationAvalanching}. Nevertheless, we still observe local fluidization of the microstructure  in the rearrangement zones resulting in the gradual collapse.

The dynamics of the observed rearrangement waves (the scaling $L_2(t)\sim t^{1/2}$)   differs from the case of flowing droplet threads ($L_2(t)\sim t$) that we reported before \cite{guzowski2022DynamicSelforganizationAvalanching}. The difference can be associated with  the different  dominant dissipation mechanism: while in flowing threads it is  the inter-droplet friction, in the printed threads it is the hydrodynamic  friction at the substrate. 

In terms of applications, we propose that the cell-encapsulating droplet-crystals printed under an external aqueous environment  could conveniently serve as drug-screening assays \cite{PompanoIsmagilovAnnRevAnalChem2011}. The printed ordered droplet assays would not only provide the advantages of  open-top, channel-free assembly but also allow to combine ultra-high   throughputs \cite{XuFangAnalChem2019} with combinatorial sampling \cite{EduatiMertenNatComm2018}.

\vspace*{0.1cm}

\noindent \emph{Acknowledgements}--The authors acknowledge financial support from National Science Center under grant NCN Sonata BIS-9 (grant no 2019/34/E/ST8/00411).

\noindent
\emph{Competing interests}--The authors of this work are the authors of a European patent application for printing of droplet chains at a substrate.


\bibliography{ArticleRefsv3} 

\end{document}